\documentclass[conference]{IEEEtran}
\IEEEoverridecommandlockouts
\usepackage{tikz}
\usetikzlibrary{arrows.meta, positioning, fit}

\usepackage[T1]{fontenc}

\usepackage{amsmath}
\usepackage{amsfonts}
\usepackage{todonotes}
\usepackage{listings}
\usepackage{colortbl}
\usepackage{hhline}
\usepackage{graphicx}
\usepackage{threeparttable,tabularx}
\usepackage{booktabs}
\usepackage{makecell}
\usepackage{multirow}
\usepackage{url}                
\usepackage{array}
\usepackage{tcolorbox}
\usepackage{ragged2e}
\usepackage{pifont}
\usepackage{subcaption}
\usepackage{adjustbox}
\usepackage{algorithm}
\usepackage{algpseudocode}
\usepackage{hyperref}
\newcommand*\circled[1]{\tikz[baseline=(char.base)]{%
            \node[shape=circle,fill=blue!20,draw,inner sep=2pt] (char) {#1};}}

\usepackage{enumitem}
\newcommand{\fei}{\textcolor{black}}
\newcommand{\modified}{\textcolor{black}}
\newtcolorbox{notebox}{
    colback=gray!5!white,
    colframe=gray!50!black,
    boxrule=1pt,
    arc=4pt,
    left=6pt,
    right=6pt,
    top=6pt,
    bottom=6pt,
    fontupper=\itshape,
    title=Takeaway:
}
\def\BibTeX{{\rm B\kern-.05em{\sc i\kern-.025em b}\kern-.08em
    T\kern-.1667em\lower.7ex\hbox{E}\kern-.125emX}}
\begin{document}

\title{Unified Framework for Qualifying Security Boundary of PUFs Against Machine Learning Attacks}

\author{
    \IEEEauthorblockN{Hongming Fei\IEEEauthorrefmark{1},
                      Zilong Hu\IEEEauthorrefmark{1},
                      Prosanta Gope\IEEEauthorrefmark{2}, and
                      Biplab Sikdar\IEEEauthorrefmark{1}}
    
    \IEEEauthorblockA{\IEEEauthorrefmark{1}\textit{National University of Singapore}, Singapore\\
    Email: \{fei.hongming, huzilong\}@u.nus.edu, bsikdar@nus.edu.sg}
    
    \IEEEauthorblockA{\IEEEauthorrefmark{2}\textit{The University of Sheffield}, Sheffield, UK\\
    Email: p.gope@sheffield.ac.uk}
}

\maketitle


\begin{abstract}

Physical Unclonable Functions (PUFs) serve as lightweight, hardware-intrinsic entropy sources widely deployed in IoT security applications. However, delay-based PUFs are vulnerable to Machine Learning Attacks (MLAs), undermining their assumed unclonability. \modified{There are no valid metrics for evaluating PUF MLA resistance, but empirical modelling experiments, which lack theoretical guarantees and are highly sensitive to advances in machine learning techniques. To address the fundamental gap between PUF designs and security qualifications, this work proposes a novel, formal, and unified framework for evaluating PUF security against modelling attacks by providing security lower bounds, independent of specific attack models or learning algorithms. We mathematically characterise the adversary’s advantage in predicting responses to unseen challenges based solely on observed challenge-response pairs (CRPs), formulating the problem as a conditional probability estimation over the space of candidate PUFs.} We present our analysis on previous ``broken" PUFs, e.g., Arbiter PUFs, XOR PUFs, Feed-Forward PUFs, and for the first time compare their MLA resistance in a formal way. In addition, we evaluate the currently ``secure" CT PUF, and show its security boundary.
We demonstrate that the proposed approach systematically quantifies PUF resilience, captures subtle security differences, and provides actionable, theoretically grounded security guarantees for the practical deployment of PUFs.

\end{abstract}
\begin{IEEEkeywords}
Physical Unclonable Function (PUF), Machine-Learning Attack, Unified Evaluation Framework.
\end{IEEEkeywords}

\section{Introduction}
The rapid proliferation of Internet-of-Things (IoT) devices has brought forward stringent demands for lightweight, scalable, and hardware-rooted authentication mechanisms. Traditional cryptographic primitives, which typically rely on secure key storage and computationally expensive operations, are ill-suited for resource-constrained edge devices. In response to this challenge, \emph{Physically Unclonable Functions} (PUFs) have emerged as an attractive alternative, leveraging inherent and uncontrollable process variations introduced during manufacturing to derive device-unique responses without requiring persistent key storage.
Formally, a PUF can be viewed as a physical function $f: \mathcal{C} \rightarrow \mathcal{R}$ that maps each input challenge $c \in \mathcal{C}$ to a response $r = f(c) \in \mathcal{R}$, thereby serving as a hardware-intrinsic fingerprint. Among various categories of PUFs, \emph{strong PUFs} are particularly valuable in authentication applications due to their support for exponentially large challenge-response spaces. A prominent class of such constructions is \emph{delay-based PUFs}, including Arbiter PUFs, XOR Arbiter PUFs, Feed-Forward PUFs, Interpose PUFs, and more recent variants such as Configurable Tristate PUFs. These designs are valued for their compact implementation and high inter-device entropy.

However, the structural characteristics that make delay-based PUFs efficient also introduce fundamental vulnerabilities. In particular, the approximately linear relationship between challenges and responses renders these constructions susceptible to \emph{Machine Learning Attacks} (MLAs).\modified{ Over the past decade, a series of works~\cite{ruhrmair2010modelling, wisiol_neural_2022,wisiol2020splitting,becker2015gap,hongming2024attacking,hongming2024optimal} have shown that with only a modest number of observed challenge-response pairs (CRPs)}, adversaries can train models—such as support vector machines, logistic regression, and deep neural networks—that accurately predict PUF responses to previously unseen challenges. This arms race between PUF designers and adversarial modelling strategies has given rise to a concerning dynamic: new PUF architectures are frequently proposed, often incorporating increased non-linearity, obfuscation mechanisms, or cryptographic primitives, in an attempt to counter known MLAs. Yet these designs are typically evaluated against a limited set of existing attack techniques, and once more powerful attacks are developed, the previously claimed security guarantees quickly become obsolete. As a result, there is a growing concern within the community regarding the \emph{stability and credibility} of claimed resistance to MLAs. This uncertainty significantly hinders the adoption of PUFs in security-critical applications.

In an effort to better understand and quantify security, \modified{researchers have employed auxiliary metrics such as response bias, inter- and intra-device uniqueness, and statistical randomness tests (e.g., entropy, PCA analysis and NIST test suites).} However, these indicators often correlate poorly with actual resistance to modelling attacks. It is not uncommon for PUF designs to exhibit favourable statistical properties and pass standard randomness benchmarks, only to be successfully modelled by new machine learning techniques shortly after their publication. \modified{For example, the Interpose PUF~\cite{nguyen2018interpose} proposed in 2018 and shown to be broken~\cite{wisiol2020splitting} in 2020; Heterogeneous Feed Forward PUF~\cite{avvaru2020homogeneous} proposed in 2020 and shown to be broken~\cite{hongming2024optimal} in 2024; high-stage XOR PUFs showing strong MLA resistance but being attacked using the reliability-based attack~\cite{becker2015gap}, where the path delay measurement has passed NIST tests~\cite{martin2021chip}.} \modified{These inconsistencies highlight the lack of a unified, interpretable, and mathematically grounded framework for quantifying PUF security against MLAs. Existing analytical models remain architecture-specific and fail to offer comparable or provable guarantees.}


\color{black}
\subsection{Motivation}
\label{subsec:motivation}
\begin{figure*}[ht]
    \centering
    \includegraphics[width=0.75\linewidth]{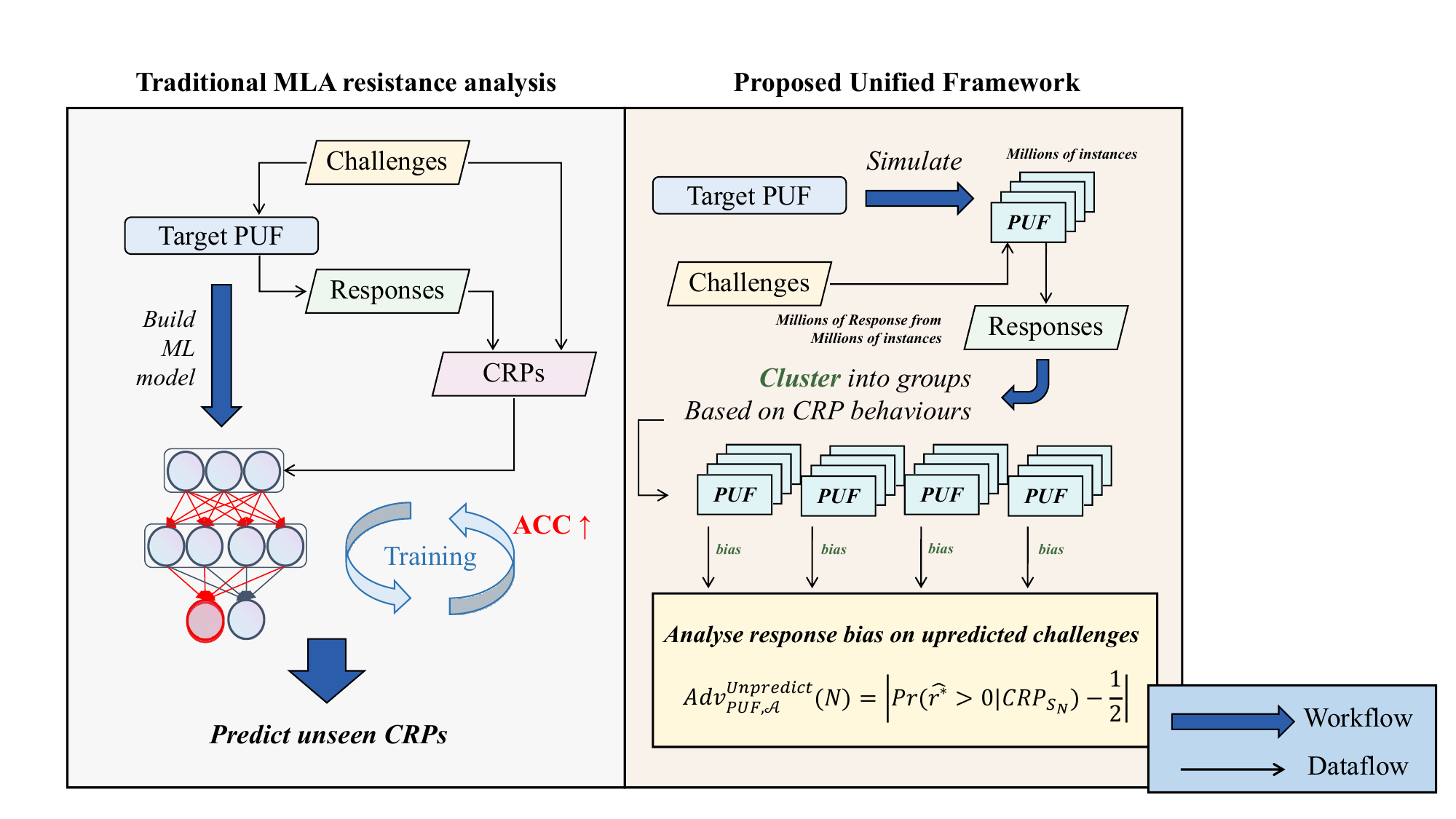}
    \caption{Comparison Between the Traditional MLA Resistance Analysis Framework and the Proposed Unified Framework.}
    \label{fig:comparison}
\end{figure*}

\color{black}
\modified{We present the traditional analysis process of MLA resistance on the left side of Fig.~\ref{fig:comparison}. First, researchers collect a set of challenge-response pairs (CRPs), then manually design and train a machine learning model tailored to the specific PUF architecture under study. The effectiveness of this approach is typically measured by the prediction accuracy on a hold-out test set. To be specific, several fundamental gaps are observed:}


\begin{itemize}
    \item \modified{\textbf{Architecture-Decoupled Evaluation:} In the traditional MLA resistance analysis methodology, different ML models are built according to different PUF architectures, resulting in fragmented and non-comparable results.}
    \item \modified{\textbf{Lack of Formal Guarantees}: Empirical results cannot give provable and formal security quantifications, hindering PUF from rigorous security applications.}
    \item \modified{\textbf{Inadequacy of Heuristic Metrics:} There is a lack of heuristic metrics which correlate well with real-world attacks. Uniqueness, bias, entropy, et al. fail to predict the MLA vulnerabilities.}
\end{itemize}

\color{black}

\subsection{Contribution}
\label{subsec:contribution}

\color{black}

In this paper, we introduce the \textbf{first formal evaluation framework} for quantifying PUF security against machine learning modelling attacks \modified{to address the above challenges. Specifically, we make the following key contributions:}

\begin{enumerate}[label=\protect\circled{\arabic*}]
    \item \modified{\textbf{A Formal Unified Framework:} We propose a novel, architecture-agnostic, and training-free framework for evaluating the resilience of PUFs against machine learning attacks. }
    
    \item \modified{\textbf{Formal Modelling with Sound Security Definitions:} We formulate a rigorous theoretical model of adversarial predictability in the form of an unpredictability game, and define a quantitative measure of adversarial advantage grounded in conditional probability theory.} 
    
    \item \textbf{Scalable and Efficient Implementation with GPU Acceleration:} \modified{We instantiate a scalable and computation-efficient implementation, powered with a Monte Carlo approximation scheme and a GPU parallelism acceleration algorithm.}
    All code and plotting scripts are open-sourced to facilitate reproducibility and community adoption.\footnote{\textbf{All codes and data used in this paper are available at:} \href{https://github.com/AnonymousAppdx/Unified-Framework-for-Defining-Security-Boundary-of-PUFs-Against-Machine-Learning-Attacks/tree/main}{\textbf{GitHub repository}}.}

    \item \textbf{Comprehensive and Actionable Security Analysis:} We conduct extensive empirical analysis under our formal framework, evaluating how parameters such as the number of stages, observed CRPs, and architectural complexity influence the adversarial advantage. These findings provide concrete guidance for designing more resilient PUF architectures and offer a principled basis for evaluating PUF-based authentication and key derivation protocols.
\end{enumerate}



\section{Related Work} \label{sec:related_work}

\textbf{\modified{Physical Unclonable Function Designs.}} Since the introduction of the Physical Unclonable Function concept by Gassend et al.~\cite{gassend2002silicon}, numerous strong PUF designs have been proposed to enable lightweight authentication. Among these, delay-based architectures, such as the Arbiter PUF (APUF)\cite{lee2004techniques}, XOR Arbiter PUFs~\cite{ruhrmair2010modelling}, and Feed-Forward Arbiter PUFs~\cite{lee2004techniques}, have been widely adopted due to their exponential challenge-response space and compact hardware footprint. However, their structural linearity and limited non-linearity have exposed them to machine learning attacks.
To enhance resilience, advanced designs such as the Interpose PUF (iPUF)\cite{nguyen2018interpose} and Configurable Tristate PUF (CT-PUF)\cite{zhang_ct_2021} have been proposed, combining multiple primitives or reconfigurable paths to obfuscate internal structures. Despite these innovations, history has shown that increased complexity often yields only transient security gains, as successive modelling techniques eventually adapt to break these constructions~\cite{ruhrmair2010modelling}.

\textbf{\modified{Machine Learning Modelling Attacks.}} Modelling attacks have evolved in tandem with strong PUF designs. Early work demonstrated that conventional supervised learning methods such as support vector machines (SVM) and logistic regression (LR) could efficiently model APUFs by exploiting their threshold-like behavior~\cite{lim2005extracting}. Subsequent studies expanded the arsenal with neural networks~\cite{ruhrmair2010modelling}, evolutionary strategies~\cite{yu_efficient_2019}, and side-channel-assisted attacks~\cite{yu_efficient_2019}.
Recent deep learning approaches~\cite{wisiol_neural_2022} have shown that even complex constructions such as 6-XOR or feed-forward PUFs can be modelled with high accuracy. Moreover, reliability-based modelling~\cite{delvaux2017machine} has proven particularly potent: by exploiting noisy but statistically informative response variations, attackers can dramatically reduce the effective learning complexity. These advances illustrate that empirical machine learning attacks have consistently outpaced ad-hoc security enhancements in PUF design.

\textbf{\modified{Theoretical Analysis and Metrics.}} While empirical studies dominate the evaluation landscape, theoretical treatments of PUF security remain comparatively scarce. Early models idealized PUFs as random oracles~\cite{gassend2002silicon,suzuki2013security}, assuming independent and uniform responses, but failed to capture realistic adversarial capabilities.
A major advance was made by van Dijk and Jin~\cite{van_dijk_theoretical_2023}, who introduced a rigorous theoretical framework that explicitly models adversaries with oracle access and partial predictive models. Their work formalized notions such as response bias, reliability, and inter-challenge correlation, providing a more nuanced and composable understanding of PUF security. However, while their framework enables provable security constructions for PUF-based key generation, it stops short of providing practical tools for evaluating PUF modelling resistance across architectures.

Entropy-based approaches offer complementary insights into PUF security by quantifying unpredictability and bias. Studies have characterized both global entropy~\cite{roel2012physically} and conditional entropy~\cite{dumoulin2024response} of PUF responses, revealing that structural bias and challenge correlation can significantly degrade effective unpredictability.
Recent works~\cite{dumoulin2024response,wang_deep_2025} have highlighted that certain challenges exhibit reduced entropy conditioned on observed CRPs, making them vulnerable to targeted modelling. However, existing analyses are often architecture-specific and rely on analytical approximations, limiting their generality across diverse PUF constructions.

\textbf{\modified{Concerns From Practical Applications.}} 
\modified{ Although PUFs are commonly used in security applications, such as authentication ~\cite{gope2018lightweight,zhang2022efficient,fei2024phenoauth} and key generation protocols~\cite{delvaux2014helper,zhang2021sc,zheng2022puf}, their resistance to machine learning attacks is typically regarded as a security assumption in the analysis. This reliance on an assumption, rather than empirical evidence, undermines the rigour of the study and raises significant security concerns.}
\section{Preliminaries}
\label{sec:pre}
In this section, we introduce the 
PUFs and their underlying mathematical logics, including Arbiter PUFs, XOR PUFs, Feed-Forward PUFs and Configurable Tristate PUFs. Then, we introduce how machine learning attacks are performed on PUFs and how the models are designed. In the end, we introduce the Monte Carlo method as the solution of statistic analysis.

\subsection{Physical Unclonable Functions and Compositions}

Delay-based Physical Unclonable Functions are the primary focus of this work due to their structural simplicity and analytical tractability. Among them, the Arbiter PUF is the most fundamental variant, and it serves as a building block for more complex compositions such as XOR PUFs, Feed-Forward PUFs, and Configurable Tristate PUFs.

\begin{figure}
    \centering
    \includegraphics[width=0.85\linewidth]{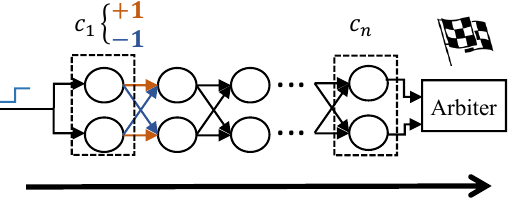}
    \caption{\modified{Delay path of an Arbiter PUF.}}
    \label{fig:apuf}
\end{figure}

\paragraph{Arbiter PUF.} 
\modified{APUF is one of the most common compositions for its simple construction and low cost~\cite{hemavathy2023arbiter,santikellur2019deep}. As shown in Figure~\ref{fig:apuf}, a}n $N$-stage APUF consists of two symmetric delay paths controlled by a challenge vector $\mathbf{c} := (c_1, c_2, ..., c_N) \in \{-1, +1\}^N$. \modified{The configuration of each stage depends on the corresponding challenge bit: if $c_i = -1$, the delay paths are crossed; otherwise, they remain parallel. An arbiter at the end determines which path signal arrives first, producing a one-bit response as $+1$ or $-1$}. 


The delay at each stage $i$ can be modelled by the difference between the two signal paths, denoted as $w_i := \delta_i^0 - \delta_i^1$. The final response depends not on the local $c_i$, but rather on the *parity* of the challenge bits from that stage onward, defined as $x_i = \prod_{j=i}^N c_j$. The complete response function can thus be written as:
\begin{equation}
    r = \mathrm{sign} \left( \sum_{i=1}^{N} w_i \cdot x_i \right).
\end{equation}


\paragraph{XOR PUF.}
To enhance non-linearity and hinder modelling attacks, Suh et al.~\cite{suh2007puf} proposed XOR Arbiter PUFs. An $n$-XOR PUF is constructed by parallelizing $n$ APUFs receiving the same challenge input. The individual responses are then XORed to produce the final output:
\begin{equation}
    r = \mathrm{sign} \left( \prod_{k=1}^{n} \sum_{i=1}^{N} w_i^{(k)} \cdot x_i \right),
\end{equation}
where $w_i^{(k)}$ denotes the delay vector of the $k$-th APUF instance. While XOR PUFs introduce combinatorial complexity, the non-linearity is still localized to the final sign function.

\paragraph{Feed-Forward PUF.}
To further increase complexity without duplicating full delay chains, the Feed-Forward PUF~\cite{lee2004technique_ff} introduces internal loops within the APUF architecture. Specifically, the intermediate output of early stages is used to influence the challenge of later stages. This introduces recursive dependencies, making modelling attacks significantly more difficult.

For a $(f_1, f_2)$-FF PUF with $n$ stages, its response can be expressed as:
\begin{equation}
\begin{aligned}
    r = \mathrm{sign} \left( \sum_{i=1}^{f_2} w_i \cdot x_i + \mathrm{sign}\left(\sum_{i=1}^{f_1} w_i \cdot x_i\right) \cdot \sum_{i=f_2}^{n} w_i \cdot x_i \right),
\end{aligned}
\end{equation}
which introduces nested sign functions, resulting in highly non-linear behavior and increased attack resistance.

\paragraph{Configurable Tristate PUF.}
Recent work proposes the Configurable Tristate PUF (CT-PUF)~\cite{zhang2021ct} as a robust architecture against MLAs by introducing both bitwise XOR obfuscation and configurable working states. CT-PUF allows for dynamic path selection and compositional complexity based on the challenge parity.

The overall logic of the CT-PUF is summarized as:
\begin{equation}
\begin{aligned}
C_{\mathrm{XOR}} &= C \oplus R_{\mathrm{Arbiter}}, \quad
R_{\mathrm{XOR}} = f_{\mathrm{CT}}(C_{\mathrm{XOR}}), \\
R_{\mathrm{Arbiter}} &= f_{\mathrm{CT}}(C_{\mathrm{Arbiter}}), \quad
R = R_{\mathrm{XOR}} \oplus R_{\mathrm{Arbiter}}.
\end{aligned}
\end{equation}

Depending on the Hamming weight and parity of specific challenge bits, $f_{\mathrm{CT}}$ is conditionally routed through one of four modes (e.g., $f_{\text{APUF}}, f_{\text{BR}}, f_{\text{RO}}$), which are formally defined as:
\begin{equation}
\small
\begin{cases}
f_{\text{APUF}}(C_{\text{even}}), & \text{if } \sum\limits_{\text{odd}} C_{\text{XOR}} = 0 \\
f_{\text{BR}}(C_{\text{OSPQO}}), & \text{if } \sum\limits_{\text{even}} C_{\text{XOR}} \bmod 2 = 1,\; \\
&\sum\limits_{\text{odd}} C_{\text{XOR}} \bmod 2 \ne 0 \\
f_{\text{BR}}(C_{\text{OQO}}) \oplus f_{\text{BR}}(C_{\text{PSP}}), & \text{if } \text{both even/odd sums} \bmod 2 = 0,\;\\
&\sum C \ne 0 \\
f_{\text{RO}}(C_{\text{OQO}}, C_{\text{PSP}}), & \text{if } \sum\limits_{\text{even}} \bmod 2 = 0,\; \sum\limits_{\text{odd}} \bmod 2 = 1 \\
\end{cases}
\end{equation}

Due to its configurability and obfuscation, CT-PUF poses significant challenges for model extraction, and its true security guarantees have remained largely empirical. In this paper, we employ our unified framework to provide the first theoretical evaluation of CT-PUFs against machine learning adversaries.


\subsection{Machine Learning Attacks on PUFs}
\color{black}
PUFs are known to be vulnerable to MLAs, as extensively demonstrated in Section~\ref{sec:related_work}. This section outlines the general process of ML-based modelling attacks on PUFs, highlighting the modelling pipeline, the attacker's capabilities, and the inherent challenges in universal evaluation. 
In what follows, we use the term \emph{black box} to emphasise that (i) the attacker typically treats the PUF as an oracle exposing only challenge--response behaviour, and (ii) the \emph{evaluation itself} provides little insight or formal guarantees beyond empirical accuracy---it does not explain \emph{why} an attack works, how the advantage scales, or what security bound holds once the attack fails.
\color{black}
As shown on the left of Figure~\ref{fig:comparison}, an MLA against a PUF follows a three-stage pipeline:

\begin{enumerate}
    \item \textbf{Data Acquisition:} The adversary collects a sufficient number of Challenge–Response Pairs (CRPs) from the target PUF. This may include the raw challenges and corresponding binary responses, configuration parameters of the PUF (e.g., number of stages, internal connections), and sometimes even auxiliary information such as side-channel leakage.

    \item \textbf{Preprocessing and Strategy Design:} The raw challenges are preprocessed based on the known structural properties of the PUF. For example, in arbiter-based PUFs, the original challenge vector $\mathbf{c}$ is typically transformed into a parity vector $\mathbf{x}$ to linearize the delay model. For composite architectures such as XOR or Interpose PUFs, attackers may employ divide-and-conquer strategies~\cite{wisiol2020splitting}, modelling subcomponents sequentially and manipulating intermediate representations to reduce model complexity. This stage also includes the selection of attack strategies, such as whether to adopt deep learning models, ensemble methods, or evolutionary techniques.

    \item \textbf{Model Construction and Training:} Based on the data and selected strategy, the attacker chooses a machine learning model (e.g., logistic regression, SVM, neural networks), tunes hyperparameters~\cite{hongming2024optimal}, and trains the model using the preprocessed CRPs. The dataset is typically split into training and testing subsets (e.g., 80\% vs. 20\%) to evaluate generalization performance.
\end{enumerate}

Despite the effectiveness of such MLAs, this attack process is highly \textit{PUF-dependent} and \textit{model-specific}. Each attack must be carefully engineered for the particular structure and logic of the target PUF. This has led to several concerns regarding the \textbf{fairness and generalizability} of security evaluations:

\begin{itemize}
    \item \textbf{PUF-specific bias in MLAs:} Classic PUFs such as APUF and XOR PUF have been extensively studied, and attack pipelines against them are mature and well-optimized. When a new PUF architecture is proposed, existing attacks may not transfer directly, leading to the \textit{false impression of improved security} merely due to the lack of compatible attack methods rather than true resilience.

    \item \textbf{Incompleteness of empirical evaluation:} The absence of known successful attacks does \textit{not imply true security}. Current MLAs are limited by the modelling techniques available and the attacker's heuristics. Security evaluations solely based on empirical resistance against selected ML models are inherently incomplete and potentially misleading.

    \item \fei{\textbf{Black-box nature and poor explainability:} In state-of-art MLA methodologies \cite{wisiol_neural_2022,hongming2024attacking,hongming2024optimal,becker2015gap}, the attacker observes only CRPs and optimizes a generic function approximator (e.g., a neural network) without a formal link to the underlying physical model. Success or failure thus depends on the chosen hypothesis class, feature engineering, and hyperparameter tuning, rather than yielding a principled bound on predictability.}
\end{itemize}

These limitations motivate the need for a \textit{theoretically grounded, model-agnostic framework} for evaluating PUF security---one that does not rely on reverse-engineering attack pipelines or model-specific assumptions. Our proposed probabilistic framework addresses this gap by offering a universal and formal method to quantify PUF vulnerability under general modelling threats, independent of any particular ML algorithm or training process.

\color{black}
\subsection{Monte Carlo Method}
The Monte Carlo method has long been recognised as a powerful technique for performing certain calculations, generally those too complicated for a more classical approach \cite{james1980monte}.
When closed-form solutions are analytically intractable—particularly in high-dimensional probability spaces—MC techniques provide scalable and statistically sound approximations, with convergence guarantees that improve with the number of samples.

In the context of PUF security evaluation, our central object of interest is the adversary’s advantage, which can be formalised as a conditional probability over a multivariate Gaussian distribution (e.g., how likely a new response can be predicted given observed challenge-response behaviour). For small-dimensional systems, analytical formulas based on correlation coefficients exist, but quickly become infeasible as the number of constraints grows. In such settings, Monte Carlo sampling offers a natural and efficient solution to approximate these conditional probabilities without requiring closed-form expressions. More details will be explained in Section~\ref{sec:meth}.
Compared to traditional machine learning attacks, Monte Carlo estimation introduces several conceptual and practical advantages. It requires no model training, no architectural assumptions, and no hyperparameter tuning. Instead of approximating the internal structure of a PUF, it focuses purely on observable input-output behaviour under a given challenge set. This paradigm shift allows for a general and architecture-independent security analysis, grounded in statistical estimation rather than attack-specific modelling.

\color{black}

\section{Methodology}
\label{sec:meth}
\begin{figure*}[ht]
\centering
\begin{adjustbox}{max width=0.7\textwidth, max height=0.7\textheight}
\begin{tikzpicture}[
    font=\sffamily,
    node distance = 3.8cm and 1.5cm,
    box/.style = {
        draw, rounded corners, minimum width=3.6cm, minimum height=3.5cm
    },
    smallbox/.style = {
        draw, rounded corners, minimum width=2.3cm, minimum height=1.6cm
    },
    labelbox/.style = {
        font=\bfseries\small, anchor=north west
    },
    arrow/.style = {-{Latex}, thick},
    every node/.append style = {font=\small}
]

\node[smallbox] (puf) at (0,0) {};
\node[box, right=of puf] (challenger) {};

\node[smallbox, right=2.3cm of challenger] (adversary) {};

\node[labelbox] at ([xshift=0.1cm, yshift=-0.1cm]puf.north west) {PUF};
\node[labelbox] at ([xshift=0.1cm, yshift=-0.1cm]challenger.north west) {$\mathcal{C}$};
\node[labelbox] at ([xshift=0.1cm, yshift=-0.1cm]adversary.north west) {$\mathcal{A}$};

\draw[arrow] ([yshift=0.3cm]challenger.west) -- node[above] {$c_i,\;c^*$} ([yshift=0.3cm]puf.east);
\draw[arrow] ([yshift=-0.3cm]puf.east) -- node[below] {$r_i,\;r^*$} ([yshift=-0.3cm]challenger.west);

\draw[arrow] ([yshift=0.3cm]challenger.east) -- node[above] {$CRPs_N,\;c^*$} ([yshift=0.3cm]adversary.west);

\draw[arrow] ([yshift=-0.3cm]adversary.west) -- node[below] {$\hat{r}^*$} ([yshift=-0.3cm]challenger.east);

\node[align=left, inner sep=2pt] (steps) at (challenger.center) {
    \begin{minipage}{3.2cm}
    ~  \\[2pt]
    ~  \\[2pt]
    $CRPs_N \leftarrow \{c_i, r_i\}_{1..N}$\\[2pt]
    $c^* \leftarrow \{0,1\}^n$\\[2pt]
    $r^* \leftarrow \mathcal{PUF}(c^*)$\\[2pt]
    Send $CRPs_N,\;c^*$ to $\mathcal{A}$\\[2pt]
    Receive $\hat{r}^*$\\[2pt]
    Check $r^* \stackrel{?}{=} \hat{r}^*$
    \end{minipage}
};

\end{tikzpicture}
\end{adjustbox}
\caption{Unseen CRP Prediction Game $\mathbf{Exp}_{\mathbf{\mathcal{PUF},\mathcal{A}}}^{Unpredict}(\mathrm{N})$.}
\label{game:prediction}
\end{figure*}
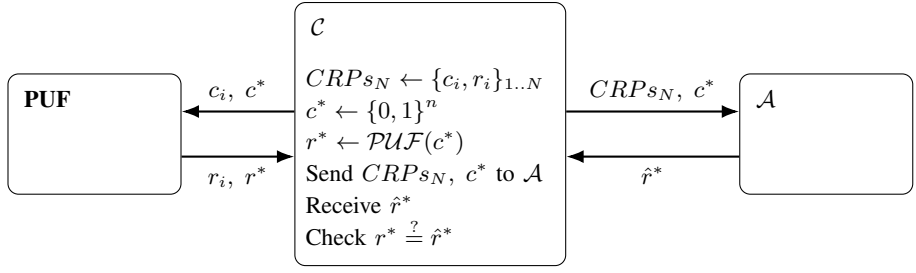


In this section, we first give the formal adversary model under the condition of MLAs. Then we define the advantage of the adversary conducting the MLAs. Based on the formal analysis, we propose a unified framework for evaluating the PUF security against modelling attack, to overcome the reliance on the ``black-box" evaluation tool, machine learning techniques and eliminate the bias involved in the evaluation. For the proposed framework, we target two research questions (RQ) to answer:
\begin{itemize}
    \item \emph{\textbf{RQ1:} How to transform the uncertainty of black-box modelling attacks into an interpretable, mathematically grounded quantification of PUF security?}
    \item \emph{\textbf{RQ2:} How to establish a rigorous, numeric security bound for PUFs based on such probabilistic quantification, independent of machine learning techniques?}
\end{itemize}

\subsection{Adversary Model}
To answer \emph{\textbf{RQ1}}, we define ``Unpredictability" game here to evaluate the PUF security against MLAs between the challenge $\mathcal{C}$ and the adversary $\mathcal{A}$ referring to \cite{armknecht2016towards}. As shown in Figure~\ref{game:prediction}, $\mathcal{C}$ holds the PUF instance $\mathcal{PUF}$ and $\mathcal{A}$ cannot have physical access to the PUF.

$\mathbf{Exp}_{\mathbf{\mathcal{PUF},\mathcal{A}}}^{Unpredict}(\mathrm{N})$:
\begin{enumerate}
    \item $\mathcal{C}$ invokes $\mathcal{PUF}$ and gets the CRP dataset $CRPs_\mathrm{{N}}$ which contains $\mathrm{N}$ challenges and responses.
    \item $\mathcal{C}$ randomly generates the challenge: $c \xleftarrow[]{Random} \{0,1\}^{n}$.
    \item $\mathcal{C}$ get the response of $c^*$ from $\mathcal{PUF}$, note as $r^*$.
    \item $\mathcal{C}$ sends $\mathrm{CRPs_{N}}$ and $c^*$ to $\mathcal{A}$.
    \item $\mathcal{A}$ tries to model $\mathcal{PUF}$ using $CRPs_\mathrm{{N}}$, and outputs his prediction of the response $\widehat{r^*}$.
\end{enumerate}

The advantage of the adversary breaks the game $\mathbf{Exp}_{\mathbf{\mathcal{PUF},\mathcal{A}}}^{Unpredict}(\mathrm{N})$ is $|Pr(r^*=\widehat{r^*})-\frac{1}{2}|$. The adversary wins the ``Unpredictability" game if $r^*=\widehat{r^*}$ holds with probability more than 1/2.

Different from \cite{armknecht2016towards,roel2012physically,armknecht2011formalization}, where the game and analysis are defined and presented from the cryptographic perspective, we define the game according to the way MLAs are performed by the MLA adversary. And the advantage that $\mathcal{A}$ wins the game is actually the accuracy in practical MLA experiments. Thus $\mathbf{Exp}_{\mathbf{\mathcal{PUF},\mathcal{A}}}^{Unpredict}(\mathrm{N})$ is more suitable for the analysis involving the ``black-box" tools, machine learning models. Furthermore, compared with getting such advantage/accuracy from the experiments, getting the metric value from performing the security game is more sound and generic for security analysis. In the following sections, we will present the unified framework on how to calculate the exact value of such advantage $\mathbf{Exp}_{\mathbf{\mathcal{PUF},\mathcal{A}}}^{Unpredict}(\mathrm{N})$, which means we are capable of evaluating the PUF security numerically and comparing them.

\subsection{A Simple Example}
\label{subsec:simple_case}
Here, we present a simple example to show how to perform the ``Unpredictable" game and how to calculate the advantage of the adversary. The example is a conditional probability problem involving Gaussian random variables, which can be easily extended to more complicated issues and solved using numerical methods.

We use the Arbiter PUF as an example, and for better understanding and explainability, the number of stages is 3. The scale of the ``training" dataset is 3, corresponding to the $\mathrm{N}$ in $\mathbf{Exp}_{\mathbf{\mathcal{PUF},\mathcal{A}}}^{Unpredict}(\mathrm{N})$. Specifically, we set $CRPs_\mathrm{{N}}:\{c_1|r_1,c_2|r_2,c_3|r_3\}$ and the target CRP $\{c^*|r^*\}$ to be:
\begin{equation}
\begin{aligned}
    &c_1:\{+1,+1,+1\};\quad r_1:+1\\
    &c_2:\{+1,+1,-1\};\quad r_2:+1\\
    &c_3:\{+1,-1,+1\};\quad r_3:+1\\
    &c^*:\{+1,-1,-1\};\quad r^*:+1 \text{ (unknown to $\mathcal{A}$)}
\end{aligned}
\end{equation}
Any $CRPs_\mathrm{{N}}$ could be chosen, as long as it does not conflict with itself (Two CRPs have opposite values but the responses are the same).
Given $CRPs_\mathrm{{N}}:\{c_1|r_1,c_2|r_2,c_3|r_3\}$ and the target challenge $c^*$, we can formulate the advantage:

\begin{align}
\label{eq:easy_case1}
&\mathbf{Adv}_{\mathbf{\mathcal{PUF},\mathcal{A}}}^{Unpredict}(\mathrm{3})\\=&\left|Pr(\text{Predicting }r^*\text{ correctly}\mid r_1>0,r_2>0,r_3>0)-\frac{1}{2}\right|\\
    =&\left|Pr(\widehat{r^*}=r^*|r_1>0,r_2>0,r_3>0)-\frac{1}{2}\right|\label{eq:e1}\\
    =&\left|Pr(\widehat{r^*}>0|r_1>0,r_2>0,r_3>0)-\frac{1}{2}\right| \label{eq:e2}
\end{align}

From Equation \ref{eq:e1} to Equation \ref{eq:e2}, it can be easily inferred that no matter whether $r^*$ is $+1$ or $-1$, the formulation remains the same since the absolute value calculation.

The mathematical formulation of the Arbiter PUF has been introduced in Section \ref{sec:pre}, thus, we can further calculate:
\begin{equation}
\label{eq:adv3}
\begin{aligned}
    \mathbf{Adv}&_{\mathbf{\mathcal{PUF},\mathcal{A}}}^{Unpredict}(\mathrm{3})=\\ Pr(&w_{1}-w_{2}-w_{3}>0\mid   w_{1}+w_{2}+w_{3}>0,\\&w_{1}+w_{2}-w_{3}>0,w_{1}-w_{2}+w_{3}>0)
\end{aligned}
\end{equation}

For an ideal Arbiter PUF, we can view \(w_1, w_2, w_3\) be three mutually independent, standard normal random variables that:
\[
w_1, w_2, w_3 \sim N(0,1), \quad \text{independent}.
\]

Now the problem is converted into the calculation of the condition probability, that the known responses $r_1>0,r_2>0,r_3>0$ are the conditions and $\widehat{r^*}>0$ is the target event. We represent the final delay difference of the three challenges in $CRP_\mathrm{3}$ to be $\Delta_1,\Delta_2,\Delta_3$ and the targe delay difference to be $d^*$ ($r=\mathrm{sign(\Delta)}$). In addition, we introduce new variables:
\[
X = w_1 + w_2,\quad Y = w_1 - w_2,\quad Z = w_3.
\]

Then the four inequalities can be rewritten:
\begin{equation}
\begin{aligned}
&\Delta_1 > 0 \;\leftrightarrow\; w_1+ w_2+ w_3> 0 
   \;\leftrightarrow\;  X + Z > 0 \;\leftrightarrow\; Z > -X.\\
&\Delta_2 > 0 \;\leftrightarrow\; w_1+ w_2- w_3> 0 
   \;\leftrightarrow\; X - Z > 0 \leftrightarrow\; Z < X.\\
&\Delta_3 > 0 \;\leftrightarrow\; w_1- w_2+ w_3> 0
   \;\leftrightarrow\; Y + Z > 0 \leftrightarrow\; Z > -Y.\\
&\widehat{\Delta^*} > 0 \;\leftrightarrow\; w_1- w_2- w_3> 0 
   \;\leftrightarrow\; Y - Z > 0 \leftrightarrow\; Z < Y.
\end{aligned}
\end{equation}

Hence, knowing the condition event
\[
\Delta_1 > 0,\; \Delta_2 > 0,\; \Delta_3 > 0
\]
is equivalent to
\[
Z > -X,\quad Z < X,\quad Z > -Y,
\]
i.e.,
\[
Z \in \bigl(\max(-X, -Y),\, X\bigr),
\]

Under this condition, we want
\[
P\bigl(Z < Y \;\mid\; Z \in (\max(-X,-Y),\, X)\bigr),
\]
then average over the distribution of \((X,Y)\).

Since \(w_1,w_2,w_3\) are mutually independent \(N(0,1)\), it follows that
\[
X = w_1 + w_2 \sim N(0,2),\quad Y = w_1 - w_2 \sim N(0,2),
\]
and \(X\) and \(Y\) are independent. Also, \(Z = w_3 \sim N(0,1)\) is independent of \((X,Y)\).

Thus, conditioned on \((X,Y)=(x,y)\), the random variable \(Z\) remains \(N(0,1)\), independent of \((x,y)\). Next, we can calculate the result and move on from Equation \ref{eq:adv3}.
\begin{equation}
\begin{aligned}
    &\mathbf{Adv}_{\mathbf{\mathcal{PUF},\mathcal{A}}}^{Unpredict}(\mathrm{3})\\&=Pr(r^*>0\mid r_1>0,r_2>0,r_3>0)\\
    &=\frac{Pr(r^*>0,r_1>0,r_2>0,r_3>0)}{Pr(r_1>0,r_2>0,r_3>0)}\\
    &=\frac{E_{X,Y}\bigl[P(Z \in (\max(-X,-Y),X)\cap\{Z < Y\})\bigr]}
        {E_{X,Y}\bigl[P(Z \in (\max(-X,-Y),X))\bigr]}.
\end{aligned}
\end{equation}

Note \(\Phi\) as the standard normal CDF.
\begin{equation}
\begin{aligned}
&P\bigl(Z \in (\max(-x,-y),\, x)\,\cap\, \{Z < y\}\bigr)\\
&= P\bigl(Z \in (\max(-x,-y),\, \min(x,y))\bigr)\\
&= \Phi\bigl(\min(x,y)\bigr) - \Phi\bigl(\max(-x,-y)\bigr).
\end{aligned}
\end{equation}
And:
\[
P\bigl(Z \in (\max(-x,-y),\,x)\bigr)
= \Phi(x) - \Phi\bigl(\max(-x,-y)\bigr).
\]
Then we integrate over \((x,y)\) in the region where \(\max(-x,-y) < x\). Recall
\[
X \sim N(0,2),\quad Y \sim N(0,2),\quad \text{independently,}
\]
so their joint density is
\[
f_{X,Y}(x,y) = f_X(x)\, f_Y(y)
= \biggl(\frac{1}{\sqrt{4\pi}} e^{-x^2/4}\biggr)
  \cdot
  \biggl(\frac{1}{\sqrt{4\pi}} e^{-y^2/4}\biggr).
\]

Hence,
\[
\begin{aligned}
&P(z>0 \mid x>0,y>0,w>0)\\
&=
\frac{\displaystyle
\iint_{\{\max(-x,-y) < x\}} 
\bigl[\Phi(\min(x,y)) - \Phi(\max(-x,-y))\bigr]
\,f_{X,Y}}
{\displaystyle
\iint_{\{\max(-x,-y) < x\}} 
\bigl[\Phi(x) - \Phi(\max(-x,-y))\bigr]
\,f_{X,Y}}.
\end{aligned}
\]

It is easy to calculate the above final equation and get the numeric results, 0.72, which means know $CRP_{\mathrm{3}}$, the accuracy of predicting the response for $c^*$ of an Arbiter PUF is 72\%. Thus the advantage that $\mathcal{A}$ can achieve is $0.72-0.5=0.22$, which is not non-trival. We can conclude that 3-stage Arbiter PUF is not secure when three CRPs are obtained by the adversary.

\begin{figure*}
    \centering
    \includegraphics[width=0.75\linewidth]{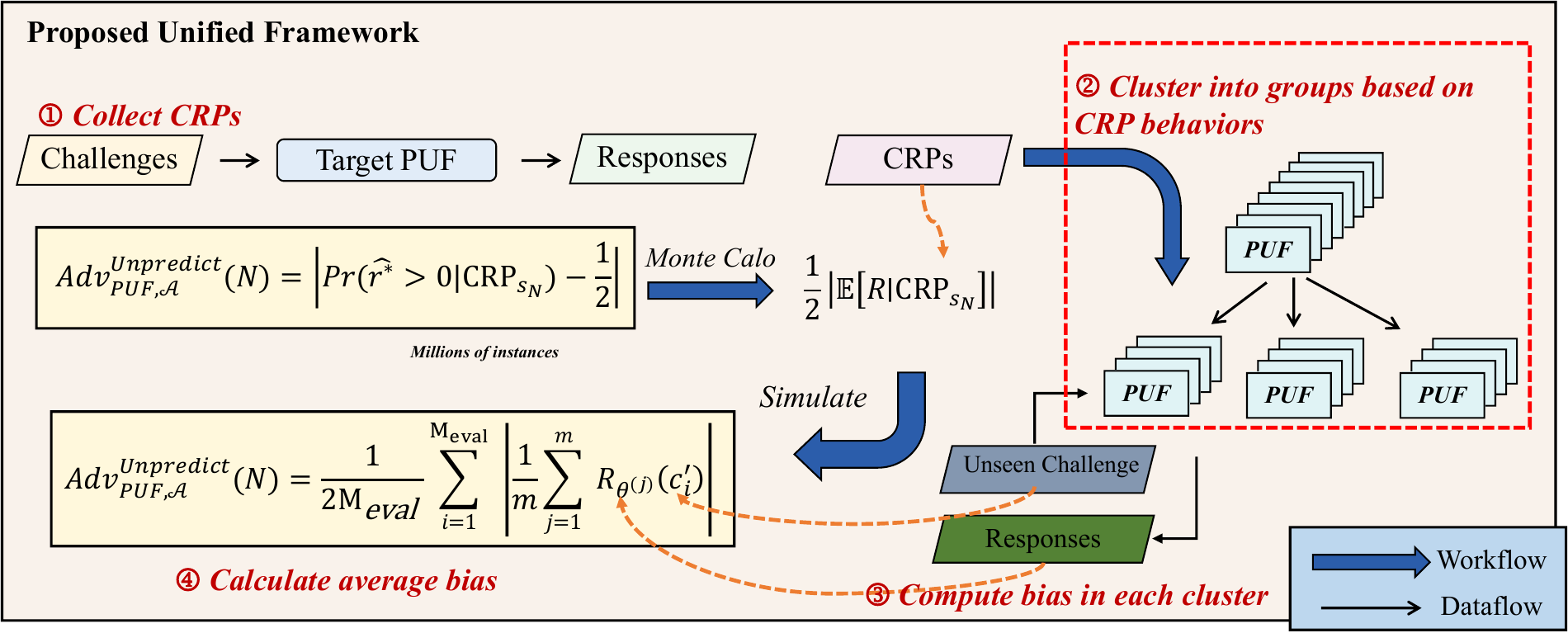}
    \caption{Proposed Unified Framework Analysing $\mathbf{Exp}_{\mathbf{\mathcal{PUF},\mathcal{A}}}^{Unpredict}(\mathrm{N})$.}
    \label{fig:method}
\end{figure*}

\subsection{Proposed Unified Framwork for Evaluating PUF Security Against MLAs}
In Section \ref{subsec:simple_case}, a simple example has been introduced where only four inequations are given as the condition, which can be viewed as the observed CRPs. Now we consider the more complicated case, with larger $\mathrm{N}$ and PUFs other than Arbiter PUFs to answer \emph{\textbf{RQ2}}. As shown in Figure~\ref{fig:method}, the detailed workflow is described.

\subsubsection{Problem Description}

Here, we give the general description for calculating the advantage $\mathbf{Adv}_{\mathbf{\mathcal{PUF},\mathcal{A}}}^{Unpredict}(\mathrm{N})$. Here, the $\mathcal{PUF}$ can be any types of PUFs as long as it follows the PUF behaviours. We have removed the limitation of the analyzing how the PUF work internally.

Given $CRPs_\mathrm{{N}}:\{c_i|r_i\}^{\mathrm{N}}$ and the target challenge $c^*$, we can formulate the advantage:

\begin{align}
\label{eq:easy_case}
\mathbf{Adv}_{\mathbf{\mathcal{PUF},\mathcal{A}}}^{Unpredict}(\mathrm{N})=\left|Pr(\widehat{r^*}>0\mid CRPs_\mathrm{{N}})-\frac{1}{2}\right| 
\end{align}

However, when $\mathrm{N}$ is smaller than 4, it is difficult to calculate the result of Equation \ref{eq:easy_case} since there does not exist general closed-form solution for calculating the probability for multivariate normal distributions. In particular, for the bivariate case where two standard normal variables $X$ and $Y$ with correlation coefficient $\rho_{xy}$ are considered, the probability that both variables are positive can be computed in closed-form as
\begin{equation}
\mathbb{P}(X > 0, Y > 0) = \frac{1}{4} + \frac{1}{2\pi} \arcsin(\rho_{xy}).
\end{equation}
Similarly, for the trivariate case involving three standard normal variables $X$, $Y$, and $Z$ with pairwise correlation coefficients $\rho_{xy}$, $\rho_{xz}$, and $\rho_{yz}$, the probability that all three variables are positive is given by

\begin{align}
&\mathbb{P}(X > 0, Y > 0, Z > 0) \\=& \frac{1}{8} + \frac{1}{4\pi} \left( \arcsin(\rho_{xy}) + \arcsin(\rho_{xz}) + \arcsin(\rho_{yz}) \right).
\end{align}

These results are derived from the properties of multivariate normal distributions and the rotational symmetry of the Gaussian measure. However, for $\mathrm{N} \geq 4$, we can not get the result directly, and numerical approximations or asymptotic methods must be employed.

\subsubsection{Monte Carlo-Based Evaluation Scheme}

To address the challenge that direct analytical computation of Equation~\ref{eq:easy_case} is infeasible for larger $\mathrm{N}$, we propose a statistical, sampling-based approach inspired by Monte Carlo methods, as shown in Algorithm~\ref{alg:puf-eval}. The core idea is to approximate the conditional probability $Pr(\widehat{r^*}>0\mid CRPs_\mathrm{{N}})$ through repeated random sampling of synthetic PUF instances and empirical evaluation.
First, Equation~(\ref{eq:easy_case}) can be rewritten as
\begin{align}
\label{eq:adv_to_mean}
\mathbf{Adv}^{\mathrm{Unpredict}}_{\mathcal{PUF},\mathcal{A}}(N)
&= \left| \Pr(R=+1 \mid \mathrm{CRPs}_N) - \tfrac{1}{2} \right|\\\
&= \frac{1}{2}\left| \mathbb{E}[R \mid \mathrm{CRPs}_N] \right|.
\end{align}
Hence, computing Eq.~(\ref{eq:easy_case}) amounts to estimating the conditional mean $\mathbb{E}[R \mid \mathrm{CRPs}_N]$.

Let $\Theta$ be the parameter space of candidate PUFs (e.g., delay vectors), and define
\[
\mathcal{S}_N \overset{\triangle}{=} \Big\{ \theta \in \Theta \; \big| \; f_\theta(c_i)=r_i,\; i=1,\dots,N \Big\},
\]
the subset of parameters that are \emph{consistent} with the observed $\mathrm{CRPs}_N = \{(c_i,r_i)\}_{i=1}^N$. If we could draw i.i.d.\ samples $\theta^{(1)},\dots,\theta^{(m)} \sim \mathcal{S}_N$, then for any new challenge $c^*$,
\begin{align}
\label{eq:mc_single}
\widehat{\mathbb{E}}[R(c^*) \mid \mathrm{CRPs}_N]
&= \frac{1}{m}\sum_{j=1}^{m} R_{\theta^{(j)}}(c^*)\\
\widehat{\mathbf{Adv}}^{\mathrm{Unpredict}}_{\mathcal{PUF},\mathcal{A}}(N; c^*)
&= \frac{1}{2}\left|\frac{1}{m}\sum_{j=1}^{m} R_{\theta^{(j)}}(c^*)\right|.
\end{align}
Averaging over $M_{\text{eval}}$ unseen challenges $\{c'_1,\ldots,c'_{M_{\text{eval}}}\}$ yields the Monte Carlo estimator
\begin{align}
\label{eq:mc_multi}
\widehat{\mathbf{Adv}}^{\mathrm{Unpredict}}_{\mathcal{PUF},\mathcal{A}}(N)
= \frac{1}{2M_{\text{eval}}}
\sum_{i=1}^{M_{\text{eval}}}
\left|
\frac{1}{m}
\sum_{j=1}^{m} R_{\theta^{(j)}}(c'_i)
\right|.
\end{align}


The above estimators are standard Monte Carlo estimators: as $m \to \infty$ (or $N_{\text{PUF}} \to \infty$ with $|G_t| \to \infty$), 
$\widehat{\mathbf{Adv}}^{\mathrm{Unpredict}}_{\mathcal{PUF},\mathcal{A}}(N)$ converges to the true value at the canonical $O(1/\sqrt{m})$ (or $O(1/\sqrt{N_{\text{PUF}}})$) rate. In our experiments, $N_{\text{PUF}}\ge 10^6$ and $M_{\text{eval}}=10^3$ keep the standard error below $10^{-3}$, rendering sampling noise negligible compared to the inter-architecture differences we report.

\color{black}

\begin{algorithm}
\caption{Monte Carlo Evaluation of PUF Predictability under Known CRPs}
\label{alg:puf-eval}
\begin{algorithmic}[1]
\Require Number of PUFs $\mathrm{N}_{PUF}$, number of known challenges $\mathrm{M}$, number of evaluation challenges $M_{\text{eval}}$, PUF architecture model
\Ensure Estimated advantage $\mathbf{Adv}_{\mathbf{\mathcal{PUF},\mathcal{A}}}^{Unpredict}(\mathrm{N})$

\State Generate $\mathrm{M}$ random challenge vectors $\{c_1, \dots, c_M\} \in \{-1, +1\}^k$
\State Randomly generate $\mathrm{N}_{PUF}$ PUF instances $\{\text{PUF}_1, \dots, \text{PUF}_N\}$
\For{$i = 1$ to $\mathrm{N}_{PUF}$}
    \State Compute response vectors $r_i = \text{PUF}_i(c_1, \dots, c_M)$
\EndFor
\State Group PUFs according to identical $r_i$ response vectors, forming groups $\mathcal{G}_1, \dots, \mathcal{G}_T$
\State Filter groups such that $|\mathcal{G}_t| \geq 2$ for evaluation
\State Generate $M_{\text{eval}}$ new challenge vectors $\{c_1', \dots, c_{M_{\text{eval}}}'\}$
\For{each group $\mathcal{G}_t$}
    \For{each PUF instance $\text{PUF}_j \in \mathcal{G}_t$}
        \State Evaluate responses $r_j' = \text{PUF}_j(c_1', \dots, c_{M_{\text{eval}}}')$
    \EndFor
    \State Compute the group-wise bias $b_t = \frac{1}{M_{\text{eval}}} \sum_{i=1}^{M_{\text{eval}}} \left| \frac{1}{|\mathcal{G}_t|} \sum_{j} r_{j,i}' \right|$
\EndFor
\State \Return Final advantage as the average group-wise bias: $\frac{1}{T} \sum_{t=1}^T b_t$
\end{algorithmic}
\end{algorithm}

The detailed procedure consists of the following key steps:

First, $\mathrm{M}$ random challenges $\{c_1, \dots, c_M\}$ are generated uniformly from the challenge space $\{-1,+1\}^k$ (Line 1), simulating the known CRPs obtained by the adversary. Next, $\mathrm{N}_{PUF}$ random PUF instances are synthesized according to the assumed PUF architecture (e.g., XOR-PUF, FF-XOR-PUF), by randomly initializing internal parameters such as delay weights (Line 2).

For each generated PUF instance $\text{PUF}_i$, its response vector $r_i$ to the known challenges $\{c_1, \dots, c_M\}$ is computed (Lines 3-4). The set of all $r_i$ vectors represents the PUF behaviors observable to the adversary.

We then group PUF instances into clusters $\mathcal{G}_1, \dots, \mathcal{G}_T$ based on identical response vectors $r_i$ (Line 5), meaning that devices within the same group cannot be distinguished by the adversary using only the known CRPs. Only groups with size greater than or equal to two are retained (Line 6), ensuring that statistical evaluation remains meaningful and avoiding trivial singleton groups.

Subsequently, $M_{\text{eval}}$ new random challenges $\{c_1', \dots, c_{M_{\text{eval}}}'\}$ are generated (Line 7) to evaluate the unpredictability of the PUFs under unseen challenges.

For each group $\mathcal{G}_t$, all PUF instances within the group are queried with the new evaluation challenges (Lines 8-10), yielding response sets $r_j'$ for each instance $\text{PUF}_j$. For each evaluation challenge, the average response across the group is computed and its absolute value is taken, yielding the per-challenge bias within the group. The group-wise bias $b_t$ is obtained by averaging these per-challenge biases (Line 11).

Finally, the estimated advantage $\mathbf{Adv}_{\mathbf{\mathcal{PUF},\mathcal{A}}}^{Unpredict}(\mathrm{N})$ is calculated as the average of the biases across all retained groups (Line 13), providing a unified statistical evaluation of the PUF's predictability under the given information constraint.

The full procedure is formally outlined in Algorithm~\ref{alg:puf-eval}.



\subsubsection{Discussion}

Algorithm~\ref{alg:puf-eval} offers a unified evaluation framework applicable to a broad class of PUF constructions, including standard XOR-PUFs, more complex structures such as Feed-Forward XOR-PUFs, and other emerging designs. By focusing on the empirical predictability conditioned on known responses, the framework circumvents the need for explicit modelling of the internal challenge-response mapping.
The grouping mechanism, based on identical response vectors, reflects an adversary's practical strategy of exploiting observed consistency among devices. By aggregating bias measurements across all sufficiently large groups (not just the largest group), the evaluation captures a more comprehensive and representative view of the PUF's resilience to modelling attacks.


\section{{Evaluation and Analysis}}
\label{sec:eval}

In this section, we describe the experimental setup and implementation details of our unified framework, followed by a quantitative analysis of adversarial advantage across various PUF architectures and parameters. Specifically, we try to answer the following questions through experiments:

\begin{enumerate}
    \item \textit{How can we compare different PUF architectures in terms of security?}
    \item \textit{How do design parameters—such as stage count or the number of CRPs—affect predictability?}
    \item \textit{Can our proposed Unified Framework scale to high-dimensional, structurally complex PUFs while maintaining reliability and efficiency?}
\end{enumerate}

By systematically analysing the adversary’s advantage under different conditions, we validate the effectiveness and scalability of our framework to quantify PUF security against machine learning-based modelling attacks.

\subsection{Experimental Setup}
\label{subsec:exp_setup}

Following Algorithm~\ref{alg:puf-eval}, we simulate a large number of synthetic PUF instances for each experiment. Each instance is instantiated from a mathematical model described in Section~\ref{sec:pre}. Key internal parameters—such as delay differences in Arbiter PUFs or feed-forward selection patterns in FF-XOR PUFs—are independently sampled from a standard Gaussian distribution to ensure idealized, unbiased behavior.

To approximate the target conditional probabilities, we generate millions of CRPs per PUF instance. This leads to substantial computational demands, which we address through GPU-accelerated, batched Monte Carlo simulation. All experiments are conducted on a dedicated Ubuntu 20.04 server with an NVIDIA A100 GPU (80 GB memory), enabling efficient evaluation of a wide range of PUF architectures and configurations.

For each experiment, we simulate over $10^6$ PUF instances to ensure that the standard deviation of the estimated adversarial advantage remains below $10^{-3}$—making statistical fluctuations negligible. All source code, experimental data, and metrics are publicly available.\footnote{\textbf{All codes and data used in this paper are available at:} \href{https://github.com/AnonymousAppdx/Unified-Framework-for-Defining-Security-Boundary-of-PUFs-Against-Machine-Learning-Attacks/tree/main}{\textbf{GitHub repository}}.}

\subsubsection{PUF Simulation at Scale}

Unlike traditional ML-based evaluations, which require access to labeled datasets and training procedures, our Monte Carlo-based method directly simulates the response behavior of synthetic PUFs under ideal conditions. For example, in Arbiter PUFs, we randomly generate the delay weight vector $\vec{W}_{1 \times k}$, where each element $w_i$ is drawn from a Gaussian distribution, $w_i \sim \mathcal{N}(0, 1)$. These parameters are initialized using PyTorch's built-in random generator:

\begin{lstlisting}
w = torch.randn(N_PUF, k, device="cuda")
\end{lstlisting}

This sampling process is extended to all supported PUF architectures using architecture-specific parameter models (see Section~\ref{sec:pre}). These simulated instances exhibit statistically ideal behavior, including near-zero bias and high inter-instance uniqueness.

\subsubsection{Efficiency and Parallelization}

The major computational challenge lies in the evaluation complexity of $N_{\text{PUF}} \times M$ operations, where $N_{\text{PUF}}$ is the number of simulated instances and $M$ is the number of observed CRPs. In our experiments, we set $N_{\text{PUF}} \geq 10^6$ and $M = 1000$, resulting in up to $10^9$ response evaluations per setting.

To overcome this, we leverage the highly parallelizable structure of PUF evaluation. For Arbiter PUFs, the response vector $\vec{r}$ for all instances over all challenges is computed using matrix multiplication and sign extraction $\vec{r} = \mathrm{sign}(\mathbf{W} \cdot \mathbf{X})$, where $\mathbf{W} \in \mathbb{R}^{N_{\text{PUF}} \times k}$ holds the delay weights and $\mathbf{X} \in \mathbb{R}^{k \times M}$ encodes challenge parity vectors. This operation is implemented using batched matrix multiplication on the GPU.

\subsubsection{End-to-End Simulation Pipeline}

We implement an end-to-end simulation pipeline in PyTorch that supports Arbiter, XOR, Feed-Forward XOR, and CT-PUF architectures. The pipeline includes the following stages:

\begin{itemize}
    \item \textbf{Step 1: PUF Instance Generation.} Delay parameters are sampled from Gaussian distributions; challenges are uniformly sampled from $\{-1, +1\}^k$ and stored on GPU.
    \item \textbf{Step 2: Response Evaluation.} For each PUF model, response vectors are computed in parallel using architecture-specific logic. For XOR and FF-XOR PUFs, multiple response paths are combined using element-wise operations such as sign products or masked selection. CT-PUFs apply conditional branching logic in tensor space based on parity patterns.
    \item \textbf{Step 3: Grouping and Advantage Estimation.} Instances with identical response patterns across observed CRPs are grouped. We compute the bias of each group on a separate challenge set and average the result to estimate the adversary’s advantage.
\end{itemize}

The entire pipeline is memory- and compute-optimized through chunk-wise batch processing and asynchronous data transfer. On an NVIDIA A100 GPU, simulating over $10^6$ PUFs with 1000 CRPs takes under five minutes.

This GPU-accelerated implementation provides two critical benefits: (1) It enables statistically robust Monte Carlo estimation by supporting large-scale sampling; (2) It makes our framework practically scalable to high-dimensional and structurally complex PUFs, which would otherwise be infeasible to evaluate using CPU-based or ML-based approaches.

\color{black}

\subsection{Response Bias Under Observed CRPs}
\label{subsec:bias_under_crp}
We begin our experimental evaluation with a fundamental question:  
\textit{How much information does a single observed CRP leak about other, unseen challenges?}

\begin{figure*}[ht]
    \centering
    \begin{subfigure}[b]{0.32\linewidth}
        \centering
        \includegraphics[width=\linewidth]{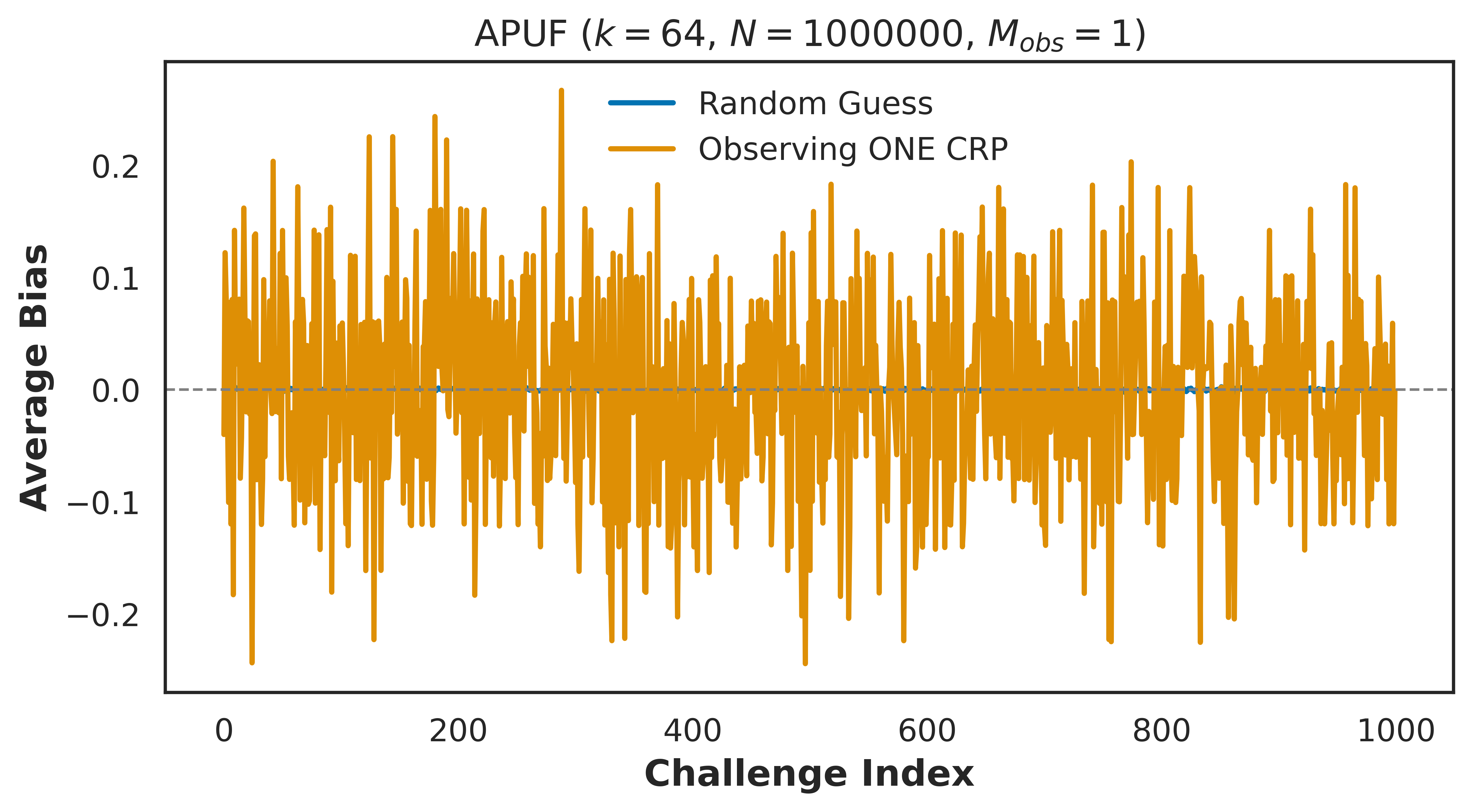}
        \caption{64-stage Arbiter PUF}
        \label{fig:apuf_1crp}
    \end{subfigure}
    \hfill
    \begin{subfigure}[b]{0.32\linewidth}
        \centering
        \includegraphics[width=\linewidth]{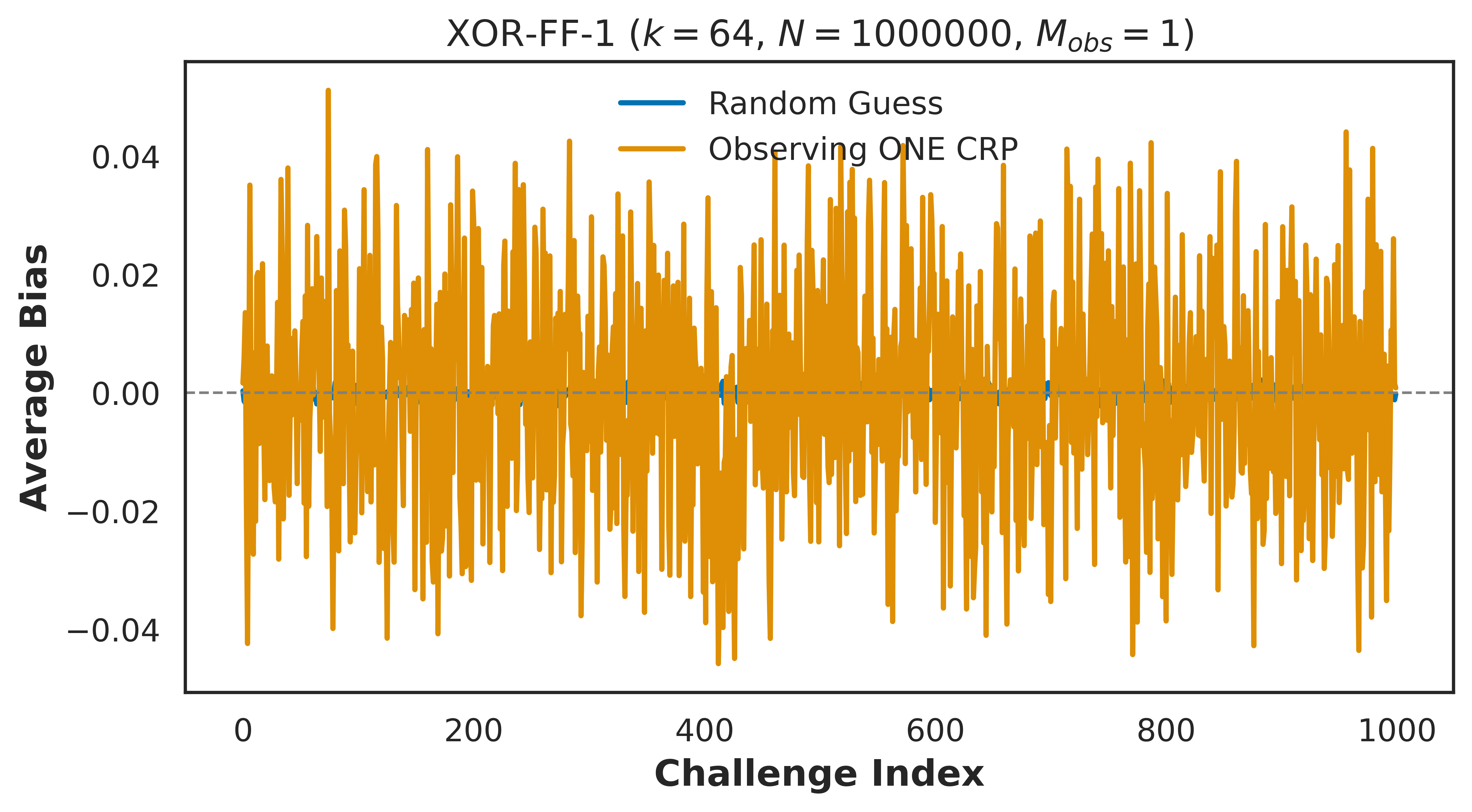}
        \caption{64-stage Feed-Forward XOR PUF}
        \label{fig:1-ffpuf_1crp}
    \end{subfigure}
    \hfill
    \begin{subfigure}[b]{0.32\linewidth}
        \centering
        \includegraphics[width=\linewidth]{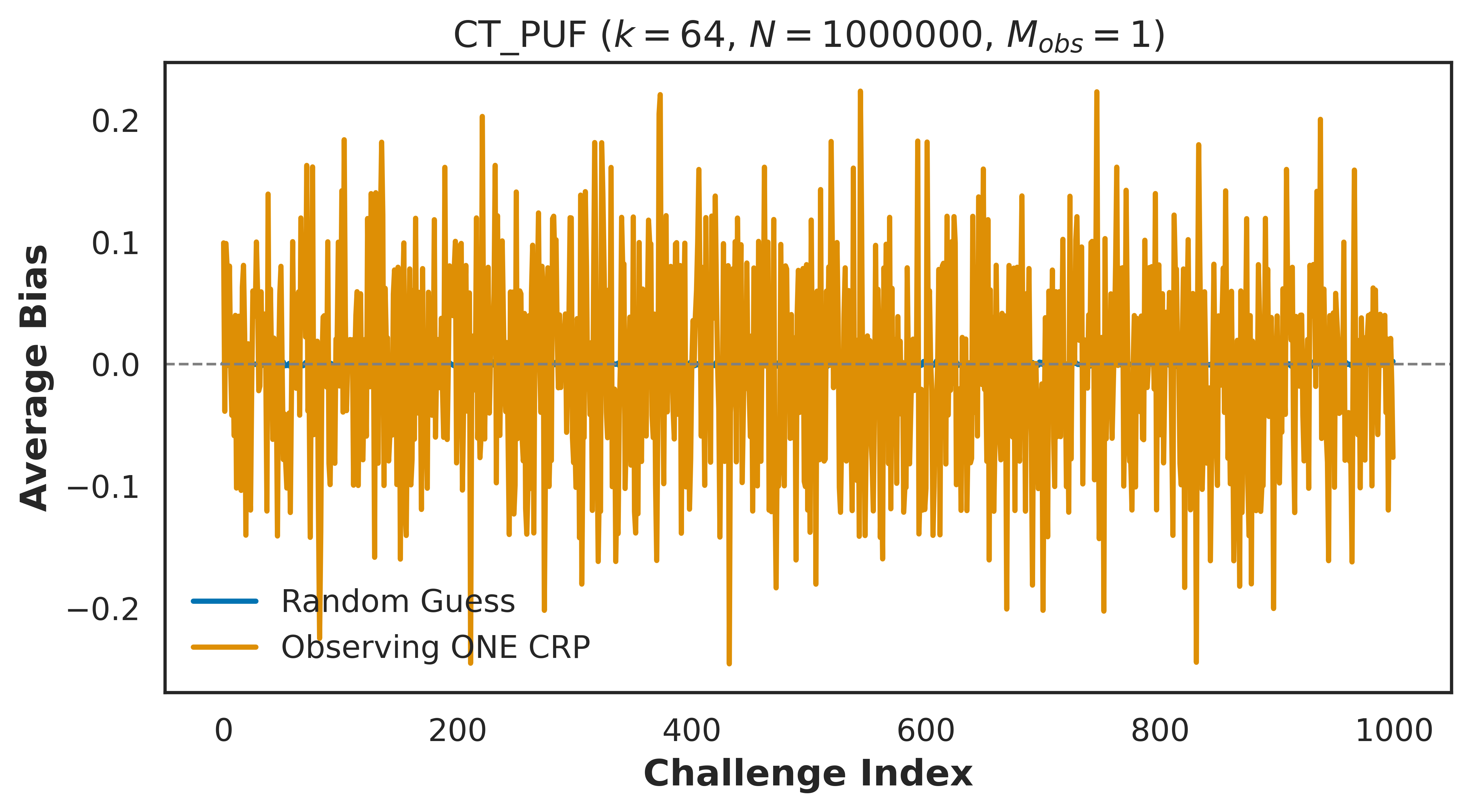}
        \caption{64-stage CT PUF}
        \label{fig:ct_c1}
    \end{subfigure}
    \caption{Response Bias of 1000 CRPs for PUFs with and without Observing One CRP}
    \label{fig:combined_bias_1crp}
\end{figure*}

This question corresponds to measuring the statistical effect of conditioning on a known CRP, as formalized in Equation~\ref{eq:easy_case}. When no CRPs are observed, the generated PUF instances behave ideally: they exhibit zero-mean response bias due to their independently sampled internal parameters.
To visualise this phenomenon, Figure~\ref{fig:combined_bias_1crp} presents a comparative study across three widely used PUF architectures. Each subfigure shows the distribution of response bias over 1000 randomly chosen challenges, both with and without conditioning on a single observed CRP. These results serve as the first step in illustrating how even minimal side information can degrade unpredictability.

\subsubsection{Evaluations on APUFs}


As shown in Figure~\ref{fig:apuf_1crp}, for 64-stage Arbiter PUFs, the distribution of mean response bias across 1000 randomly selected challenges remains centred around zero (blue line) when no CRP is conditioned (which is equivalent to random guessing), confirming the statistical ideality of the generated instances. However, upon conditioning on the observed response of a single CRP, a significant deviation from zero bias emerges. We present the 1000000 PUFs' responses on the 1000 unseen challenges and plot their bias, which is considered as the adversary's advantage of predicting the correct response. This phenomenon demonstrates that even minimal side information substantially skews the adversary's predictive capabilities.

\begin{notebox}
Observing even one single CRP introduces measurable bias in predicting unseen responses, confirming that delay-based PUFs inherently leak information once partial challenge-response knowledge is exposed.
\end{notebox}

\subsubsection{Evaluations on Feed-Forward PUFs}
Similarly, for 64-stage 1-FF-XOR PUFs shown in Figure~\ref{fig:1-ffpuf_1crp}, the unconditioned bias without observing any CRPs remains negligible, but noticeable bias appears after observing one CRP. Although the induced bias is somewhat less pronounced than that of standard Arbiter PUFs, e.g., 0.1-0.2 vs. 0.02-0.04, the feed-forward structure does not entirely eliminate the adversary's information gain. These results highlight the inherent vulnerability introduced by partial knowledge of a PUF's behaviour, even in structurally more complex designs.

\begin{notebox}
The feed-forward structure introduces non-linear obfuscation, but partial observations still allow a measurable advantage.
\end{notebox}


\color{black}
\subsubsection{Evaluations on CT PUFs}
APUFs, XOR PUFs, and Feed-Forward PUFs have already been demonstrated to be susceptible to machine learning attacks, making the observed bias in their responses to unseen challenges quite predictable. In the case of CT PUFs, to the best of our knowledge, there have been no reported successful attacks to date. Now the question arises: Is it possible to assess security using the proposed unified framework? As shown in Figure~\ref{fig:ct_c1}, a non-trivial bias exists on the unseen responses after the adversary observes one CRP, which indicates that the CT PUF is not secure. The seemingly stable status quo we have today will be disrupted in the future as carefully designed machine learning models are introduced. This insight highlights that current evaluation tools—such as randomness test suites, bias metrics, and PCA (Principal Component Analysis) tools—are inadequate for conducting thorough security evaluations.

\begin{figure}[ht]
    \centering
    \includegraphics[width=0.85\linewidth]{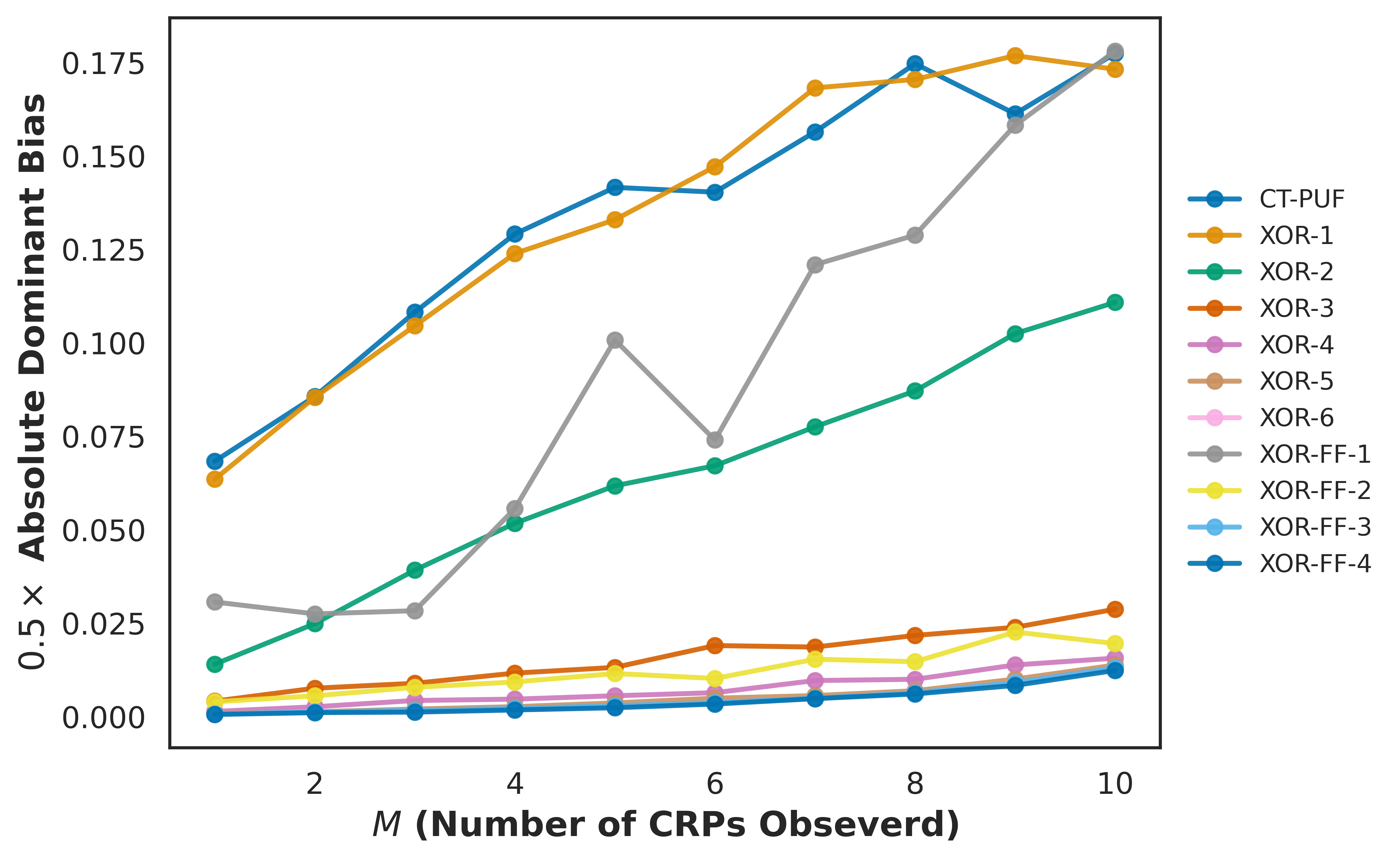}
    \caption{The Adversary Advantage of APUFs, XOR-PUFs, FF-XOR-PUFs and CT PUFs after $\mathrm{N}$ CRPs are observed.}
    \label{fig:comparison-pufs}
\end{figure}

\subsection{Impact of the Amount of CRPs}
In PUF MLA analysis, the number of CRPs is essential. Generally, a larger number of CRPs leads to improved modelling performance, such as increased accuracy. However, this relationship has not been formally evaluated or clearly explained. According to Equation~\ref{eq:easy_case}, $\mathbf{Adv}_{\mathbf{\mathcal{PUF},\mathcal{A}}}^{Unpredict}$ is related to the amount of observed CRPs, $\mathrm{N}$.
As shown in Figure~\ref{fig:comparison-pufs}, the adversary's advantage on guessing unseen responses from APUFs, XOR PUFs, FF-XOR-PUFs and CT PUFs is evaluated after $\mathrm{N}$ CRPs being observed. We can find that, overall, with larger $\mathrm{N}$, $\mathbf{Adv}_{\mathbf{\mathcal{PUF},\mathcal{A}}}^{Unpredict}$ increases. The finding aligns with experimental observations from MLAs, that increasing the amount of training data can increase the model prediction accuracy on unseen CRPs. Although it is now widely accepted that ``More training data makes stronger attacks", we have for the first time explicitly demonstrated this phenomenon. Since in practical training, the ML models have the local minimum and underfitting problems, before $\mathrm{N}$ reaches a threshold, the modelling accuracy is stuck around 50\%. With the help of the unified framework, we can make the claim that: ``More training data makes stronger attacks".

\color{black}

\subsection{Impact of the Number of Stages}
\label{subsec:impact_k}
\begin{figure}[t]
    \centering
    \includegraphics[width=0.8\linewidth]{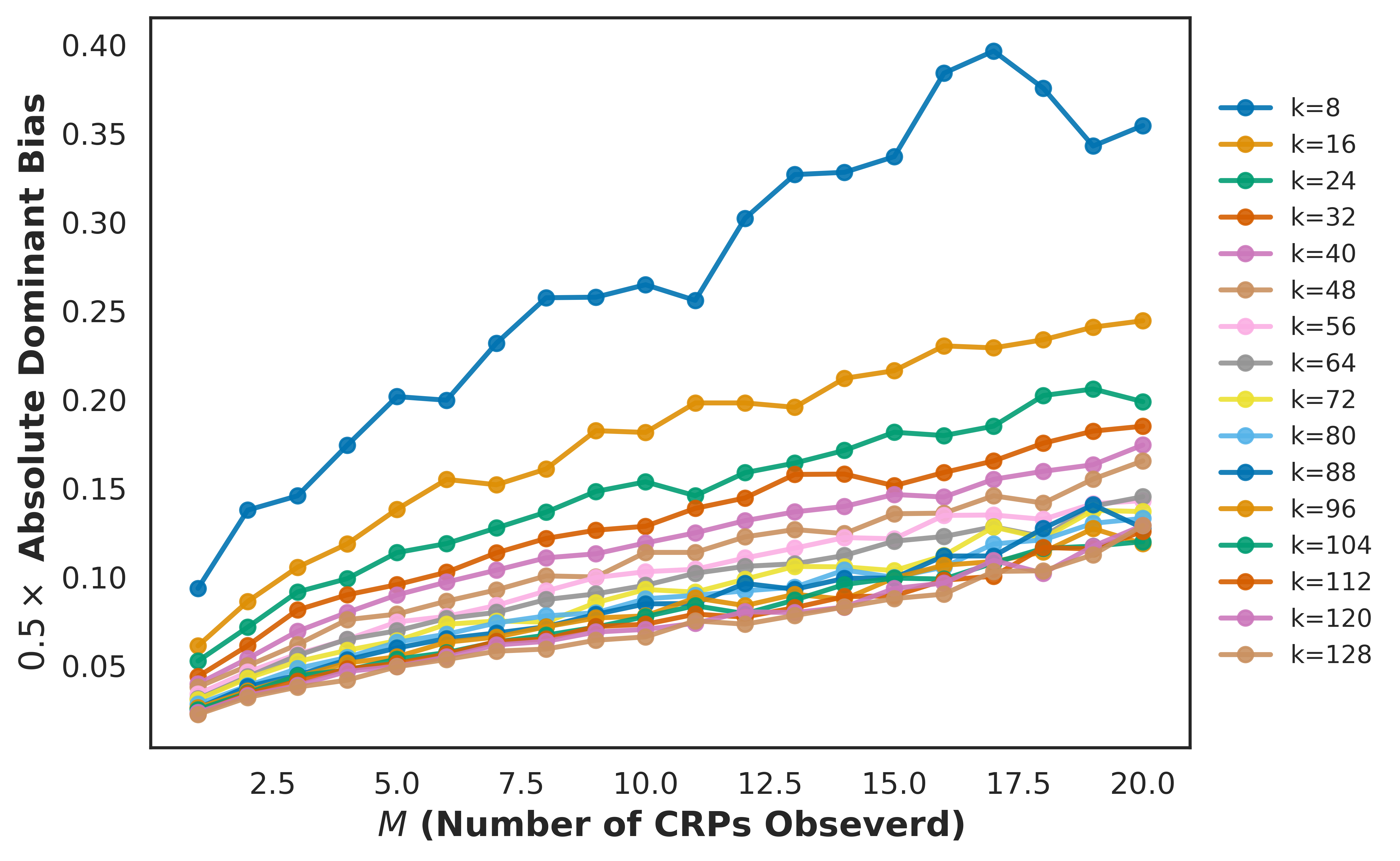}
    \caption{Adversary Advantage for Arbiter PUFs with Different Numbers of Stages $k$}
    \label{fig:apuf_k}
\end{figure}

The number of stages $k$ is important to PUF designs, considering from security and hardware implementation. Normally, it is widely accepted to set $k=64$ or $k=128$. However, there is a lack of theoretical support in setting the parameter. Is it similar to normal cryptographic primitives that, in that the security level will be improved to the next level when the secret length is beyond 256?
To analyse the influence of the number of stages $k$ on PUF security, we evaluate the adversary's advantage for APUFs with varying $k$ values. As depicted in Figure~\ref{fig:apuf_k}, the advantage consistently decreases as the number of stages increases. This behaviour is attributed to aggregating independent random delays across more stages, leading to a more unpredictable final response.

Notably, the security improvement saturates beyond $k=32$ stages. The marginal reduction in adversary advantage between $k=64$ and $k=128$ is relatively minor, indicating that simply extending the number of stages yields diminishing returns beyond a moderate threshold. This observation corroborates experimental findings in machine learning-based modelling attacks, where training accuracy plateaus as the challenge dimension increases. Therefore, while longer PUF structures provide enhanced resistance, architectural innovations are required to achieve substantial security gains beyond a certain complexity level. For increasing $k$, the hardware overhead also increases proportionally.

\begin{notebox}
Increasing the number of PUF stages $k$ improves resilience by reducing the adversary's prediction advantage. However, the marginal security gain diminishes for large $k$, indicating asymptotic saturation in practical protection.
\end{notebox}

\subsection{Comparison Across Different PUF Architectures}
\label{subsec:comparison_pufs}

We further perform a comparative evaluation across different PUF architectures under identical experimental settings to investigate the effectiveness of structural enhancements.

\begin{notebox}
    XOR-PUFs and Feed-Forward XOR-PUFs offer significant improvements in unpredictability compared to basic APUFs, due to amplified non-linearity and masking effects in the CRP mapping, which aligns with the experimental results.
\end{notebox}
Figure~\ref{fig:comparison-pufs} presents the adversary's advantage for 64-stage APUFs, XOR-PUFs, and FF-XOR-PUFs as the number of known CRPs $\mathrm{M}$ varies. It is observed that XOR-PUFs achieve substantially lower adversary advantage compared to standard APUFs, validating the intuition that XORing multiple independent Arbiter PUF outputs increases the effective entropy and obscures the modelling surface.
Furthermore, FF-XOR-PUFs exhibit even lower predictability than XOR-PUFs across the entire range of $\mathrm{M}$. The feed-forward mechanism introduces non-linear dependencies among the internal nodes, disrupting the simple linear separability that modelling attacks often exploit. This structural complexity renders the FF-XOR-PUFs significantly more resilient against both traditional machine learning and advanced inference strategies. The results clearly demonstrate that structural augmentation of PUF architectures, mainly through introducing non-linearities and internal feedback, is critical for improving resistance against modelling attacks, primarily as the adversary accumulates more observations.
\vspace{-2mm}
\subsection{MLA Benchmarks \emph{vs.} Unified-Framework Findings}
\label{subsec:mla_vs_unified}
\begin{table}[t]
\centering
\caption{\modified{Classical MLA Results}.}
\label{tab:mla-crp}
\begin{tabular}{lll}
\toprule
\textbf{PUF Type} & \textbf{\modified{Training CRPs~\cite{hongming2024attacking}}} & \textbf{\modified{Training CRPs~\cite{wisiol_neural_2022}}} \\ \midrule
APUF (64-stage)   & $2k$ & $2k$ \\ \cmidrule(l){1-3}
 2-XOR     & $8k$  & $10k$                 \\ 
 4-XOR     & $80k$  & $30k$                 \\ 
 6-XOR     & $800k$  & $2$ M               \\ 
 7-XOR     & $2.4$ M & $20$ M              \\ \cmidrule(l){1-3}
 1-FF      & $20k$ & $40k$                \\ 
 2-FF      & $90k$ & $120k$                \\ 
 3-FF      & $120k$ & $20k$               \\ \cmidrule(l){1-3}
CT-PUF           & (not broken) & (not broken)      \\ \bottomrule
\end{tabular}
\end{table}

\begin{figure}
    \centering
    \includegraphics[width=0.95\linewidth]{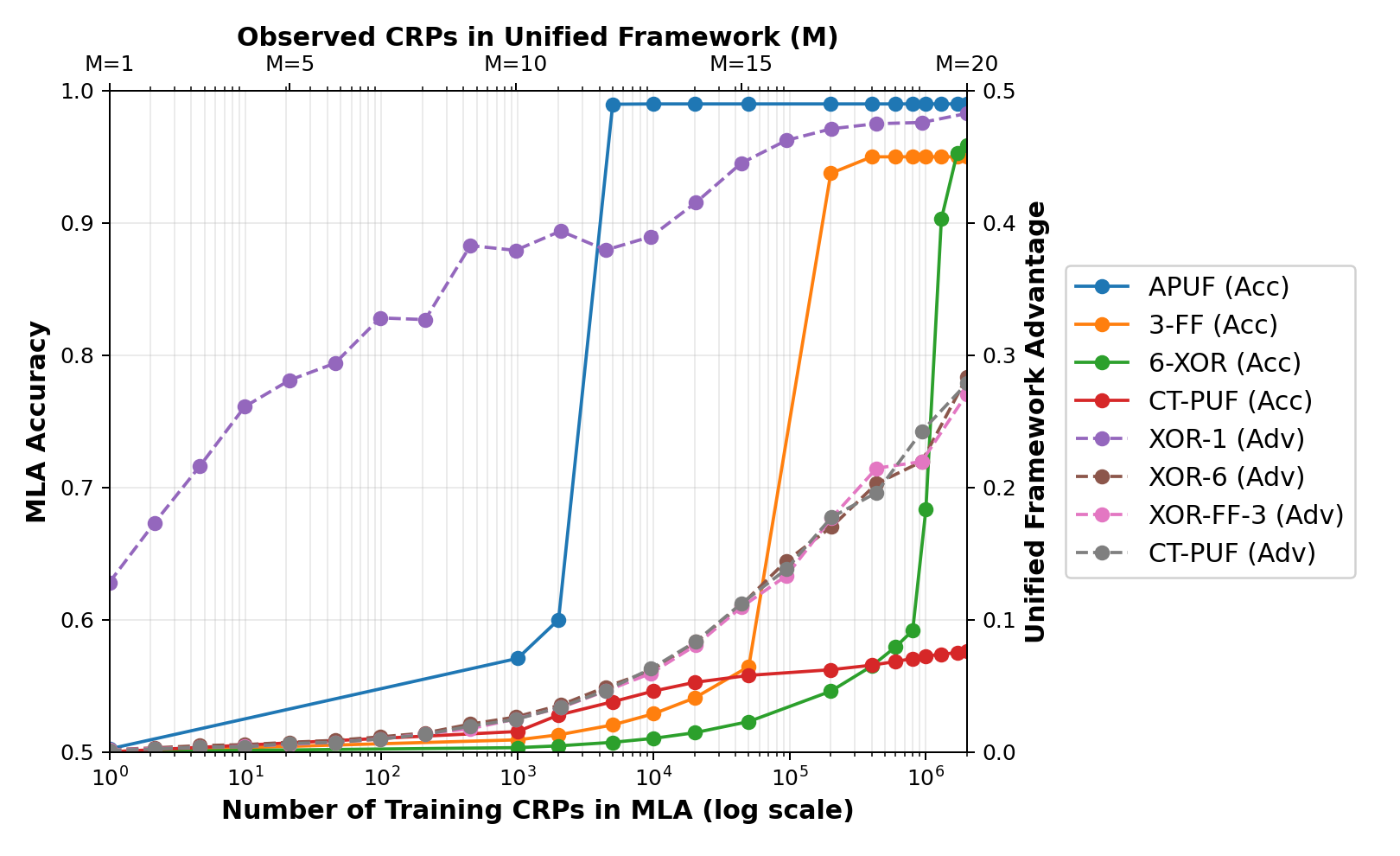}
    \caption{\modified{Classical MLA effort vs.\ unified-framework insights.}}
    \label{fig:mla_vs_unified}
\end{figure}

\begin{table*}[ht]
\centering
\caption{\modified{Comparison between traditional MLA benchmarks and the proposed unified framework.}}
\label{tab:mla-unified-comparison}
\renewcommand{\arraystretch}{1.15}
\begin{tabular}{p{3.2cm}p{6.2cm}p{6.2cm}}
\toprule
\textbf{Aspect} & \textbf{MLA Benchmarks} & \textbf{Unified Framework (This Work)} \\
\midrule
\textbf{Evaluation paradigm} & 
Empirical, model-specific; requires designing and training ML models for each PUF architecture. &
Formal, architecture-agnostic; computes adversarial advantage via probabilistic inference without ML training. \\


\textbf{Metric reported} &
Prediction accuracy (bimodal behaviour: $\approx$50\% or $>$90\%). &
Expected adversarial advantage $\mathbf{Adv}_{\mathcal{PUF},\mathcal{A}}^{Unpredict}(N)$ as a continuous function of $N$. \\

\textbf{Interpretability} &
Attack-specific; limited generalization across PUF types. &
Unified and interpretable; comparable security bounds across architectures. \\

\textbf{Security insight} &
Reactive: indicates vulnerability only \emph{after} a successful ML attack. &
Proactive: reveals bias or learnability even before an attack is developed. \\

\textbf{Computation overhead} &
High (training time grows with CRP size and model complexity). &
Low (closed-form probability estimation with parallel Monte Carlo). \\

\textbf{Design-stage usability} &
Not suitable; requires advanced MLA techniques. &
Lightweight screening tool for new architectures prior to fabrication. \\
\bottomrule
\end{tabular}
\end{table*}

To place our probabilistic results in perspective, we contrast them with the empirical MLA reported by Hongming \emph{et al.}~\cite{hongming2024attacking} \modified{and Wisol \emph{et al.}~\cite{wisiol_neural_2022}, which covered the attack benchmarks across all studied PUFs.} Table~\ref{tab:mla-crp} juxtaposes the \emph{number of CRPs} required to reach $>\!90\%$ prediction accuracy in that work with the \emph{qualitative insights} obtained from our unified framework. \modified{Figure~\ref{fig:mla_vs_unified} plots the MLA-CRPs relationship using the proposed unified framework and classical MLAs.}

\textbf{Alignment and divergence.}  For architectures already broken by MLAs (e.g.\ APUF, low-order XOR), our framework reports large adversarial advantages even at $N\!=\!1$, matching the practical ease of modelling.  As the XOR order or feed-forward depth increases, both approaches show security gains. Yet, our metric reveals a \emph{boundary effect}: beyond 5-XOR (or 3-FF) the incremental reduction in bias becomes marginal. At the same time, MLA needs exponentially more CRPs—highlighting a point of diminishing returns where added hardware cost brings limited benefit.

\textbf{Proactive insight.}  For CT-PUF, no successful MLA has been published, so classical benchmarking offers no guidance \modified{as shown in Figure~\ref{fig:mla_vs_unified}}.  Our framework, however, immediately exposes a non-trivial bias after just one observed CRP, signalling latent vulnerability and suggesting that the current “secure’’ status is merely the absence of a specialised attack model rather than intrinsic unpredictability.

\textbf{Practical implication.}  \modified{The comparisons shown in Table~\ref{tab:mla-unified-comparison} underscore two complementary roles: MLA studies quantify \emph{how quickly and accurately} an existing model can be trained, whereas our unified framework supplies an \emph{architecture-independent upper security bound} on predictability—even before an attack is devised—thus providing designers a lightweight, early-stage tool for security screening and parameter tuning.}
It is important to note that the success criterion in these modelling attacks is not simply achieving high prediction accuracy in an absolute sense. For PUFs, especially delay-based designs, the accuracy distribution in machine learning modelling typically exhibits a bimodal behaviour: when the attack fails, the prediction accuracy remains close to random guessing (around 50\%–60\%); when the attack succeeds, the accuracy quickly jumps above 90\%. Consequently, it is insufficient to use raw prediction accuracy as the sole metric for evaluating an attack's effectiveness.
\modified{In comparison, if PUF designers utilize the proposed unified framework as a metric for evaluation just like randomness and bias and so on before publishing a new PUF design, they can ``immediately" get a formal and provable security estimation. Here ``immediately" means, there is no need waiting for the development of ML techniques or the community to propose new attack strategies. For example, when the designers of CT PUF evaluate the response bias as nearly 50\%, they can confidently claim that the CT PUF is practically unbiased. The additional work is: 1) Formulate the mathematical model of the PUF, which is already done by all the PUF designers when they publish new PUFs; 2) Simulate PUF instances, in parallel if possible for efficiency; 3) Apply \textbf{Algorithm~1} and calculate average bias as the advantage $\mathbf{Adv}_{\mathbf{\mathcal{PUF},\mathcal{A}}}^{Unpredict}(\mathrm{N})$. Notably, our framework provides a definitive quantitative bound without requiring the time-consuming train–test–retrain cycles of MLA benchmarks. This enables a shift from post-hoc empirical validation to pre-emptive, theory-grounded assurance.}

\section{Conclusion}
\label{sec:conclusion}

\modified{This paper presented a formal and unified framework for evaluating the security of Physical Unclonable Functions against machine learning attacks. Unlike empirical benchmarking that depends on model-specific heuristics or attack algorithms, the proposed approach quantifies security through a probabilistic formulation of adversarial advantage—requiring no training or learning process. By estimating conditional response distributions via Monte Carlo simulation, it offers an interpretable and architecture-agnostic measure of PUF unpredictability.
Our evaluations reveal that this formalism captures subtle security distinctions among PUF architectures and exposes the latent impact of partial CRP leakage. Even a single observed CRP can induce measurable bias in unseen responses across multiple designs, including those previously believed to resist modelling attacks such as CT-PUFs. Furthermore, while structural augmentations such as XOR and feed-forward mechanisms improve robustness, their benefits saturate rapidly, highlighting the diminishing returns of purely architectural complexity. While the framework reveals the inherent limits of learnability in existing designs, it does not diminish the value of PUFs as lightweight physical primitives; rather, it enables a clearer understanding of their achievable security under realistic entropy constraints. By bridging formal analysis with empirical practice, this work establishes a principled basis for transparent, comparable, and future-proof security evaluation of PUFs, paving the way toward more trustworthy and interpretable hardware-rooted cryptographic systems.}



\bibliographystyle{IEEEtran}
\bibliography{reference}

\end{document}